\documentclass[12pt,draftcls,onecolumn,twoside]{IEEEtran}
\usepackage{subfigure}
\usepackage{colortbl}
\usepackage{bm}
\usepackage{fancyhdr}       % for page header
\usepackage{graphicx}  % Written by David Carlisle and Sebastian Rahtz
\usepackage{url}       % Written by Donald Arseneau

\usepackage{amsmath}   % From the American Mathematical Society
\usepackage{cite}

\usepackage{amsfonts,amssymb}

\usepackage{stfloats}

\usepackage{cases}
\usepackage{algorithm}
\usepackage{multirow}
\usepackage{algorithmic}
\newcommand{\ls}[1]
    {\dimen0=\fontdimen6\the\font
     \lineskip=#1\dimen0
     \advance\lineskip.5\fontdimen5\the\font
     \advance\lineskip-\dimen0
     \lineskiplimit=.9\lineskip
     \baselineskip=\lineskip
     \advance\baselineskip\dimen0
     \normallineskip\lineskip
     \normallineskiplimit\lineskiplimit
     \normalbaselineskip\baselineskip
     \ignorespaces
    }

\begin{document}
\title{{\huge Orbital Angular Momentum for \\ Wireless Communications}}
%Orbital-Angular-Momentum Based Radio Vortex Communications for Heterogenous\\Ultra-Dense Networks
\author{Wenchi Cheng,~\IEEEmembership{Member,~IEEE}, Wei Zhang,~\IEEEmembership{Fellow,~IEEE}, Haiyue Jing,~\IEEEmembership{Student Member,~IEEE}, Shanghua Gao, and Hailin Zhang,~\IEEEmembership{Member,~IEEE}\vspace{3pt}
\\
%Here I still need to add corresponding author's information to make the first paper filled.
\thanks{\ls{.5}

Wenchi Cheng, Haiyue Jing, Shanghua Gao, and Hailin Zhang are with the State Key Laboratory of Integrated
Services Networks, Xidian University, Xi'an, 710071, China (e-mails: wccheng@xidian.edu.cn; hyjing@stu.xidian.edu.cn; shgao@stu.xidian.edu.cn; hlzhang@xidian.edu.cn).

Wei Zhang is with the School of Electrical Engineering and Telecommunications, University of New South Wales Sydney, NSW, Australia (w.zhang@unsw.edu.au).

}
%\thanks{Digital Object Identifier 0000000000000000}
%\vspace{-55pt}
}
\date{\today}

%\markboth{TO BE SUBMITTED TO IEEE WIRELESS COMMUNICATIONS}{TO BE SUBMITTED TO IEEE WIRELESS COMMUNICATIONS}

\maketitle

\thispagestyle{empty}

% make the title area
%\maketitle
%\thispagestyle{empty}

%\vspace{-5pt}
\begin{abstract}
\ls{1.2}
As the traditional resources (frequency, time, space, etc.) are efficiently utilized, it becomes more and more challenging to satisfy the ever-lasting capacity-growing and users-boosting demand in wireless networks. Recently, the electromagnetic (EM) wave was found to possess not only linear momentum, but also angular momentum. The orbital angular momentum (OAM) is a kind of wavefront with helical phase. The OAM-based vortex wave has different topological charges, which are orthogonal to each other, bridging a new way for multiple access in wireless communications. In this article, we introduce the fundamental theory of OAM and the OAM based wireless communications. The research challenges regarding OAM signal generation, OAM beam converging, and OAM signal reception are discussed. Further, we propose a new multiuser access with different OAM-modes in wireless networks, where multiple OAM-modes are used as a new orthogonal dimension for interference avoidance. Simulation results reveal the inherent property of OAM waves and show that OAM based radio transmission can significantly increase the spectrum efficiency in wireless networks.
\end{abstract}

%\vspace{-20pt}
\begin{IEEEkeywords}
%\vspace{-5pt}
\ls{1.0}
Orbital angular momentum (OAM), radio vortex wireless communications, multiuser access, interference avoidance.
\end{IEEEkeywords}
%\vspace{-10pt}

\newpage

\setcounter{page}{1}

\IEEEoverridecommandlockouts

%\IEEEpeerreviewmaketitle

{%\ls{1.0}
\begin{center}
{\Large Orbital Angular Momentum for Wireless Communications}
\end{center}}
\vspace{-15pt}

\maketitle

\section{Introduction}
As wireless communications migrate from the fourth-generation (4G) to the fifth-generation (5G) and beyond, it is highly demanded to meet the requirements of explosive data traffic. For example, the aggregate data rate is expected to be increased by roughly 1000 times for 5G as compared with 4G~\cite{5G_Jeff}. It is expected using 5G and 5G-beyond New Radio (NR) for spectrum efficiency enhancement with advanced techniques, such as massive multiple-input-multiple-output (MIMO), co-frequency co-time full-duplex, and millimeter-wave (mmWave)~\cite{5G_Jeff,5G_ITU}. For 5G NR, the promise of significant spectrum efficiency enhancement, vast spatial diversity, and simple transmit/receive structure has elevated massive MIMO to a central position in 5G wireless communications networks, with a foreseen role of coexisting with mmWave~\cite{5G_five_maga}. Co-frequency co-time full-duplex, which potentially double the spectrum efficiency, is expected to be integrated into future 5G-beyond wireless communications networks~\cite{5G_app_maga}. However, during the past few decades, multiple orthogonal resources, such as frequency, time, and space, were extensively explored. Nowadays, it becomes more and more difficult to increase capacity or support more users with the traditional access techniques such as time-division-multiple-access and frequency-division-multiple-access.
%~\cite{5G_5_disr}

In fact, until now wireless communication is being built on the plane-electromagnetic (PE) wave. However, the electromagnetic (EM) wave possesses not only linear momentum, but also angular momentum, which contains the spin angular momentum (SAM) and orbital angular momentum (OAM). OAM, as a kind of wavefront with helical phase, has attracted much research attention~\cite{Thed_2007}. OAM has a great number of topological charges, i.e., OAM-modes. Beams with different OAM-modes are orthogonal to each other and they can be multiplexed/demultiplexed together, thus increasing the capacity without relying on the traditional resources such as time and frequency. OAM, which has multiple orthogonal topological charges, bridges a new way to significantly increase spectrum efficiency and is expected to be used in 5G-beyond or even more future wireless communications networks.

Recently, some experiments have verified the feasibility of OAM based wireless communications~\cite{Thed_2007,118001,6425378,LOS_OAM_2017}. The authors of~\cite{118001} studied two OAM-modes (OAM-modes $0$ and $1$), which share the same frequency band. The authors of~\cite{6425378} performed the OAM based high capacity transmission with $60$ GHz and $17$ GHz carrier frequencies, respectively. The authors of~\cite{Yan2016} demonstrated that OAM multiplexing can achieve high capacity in mmWave communications. Moreover, the research on OAM based wireless communications extends to mode detection, mode separation, axis estimation and alignment, mode modulation, OAM-beams converging, and mode hopping, etc. Generally, MIMO multiplexing can be jointly used with OAM, thus significantly increasing spectrum efficiency. The authors of~\cite{LOS_OAM_2017} and~\cite{OAM_Wenchi_2017} experimentally and theoretically demonstrated that jointly using MIMO based spatial multiplexing and OAM multiplexing can increase the spectrum efficiency of wireless communications.

Although the feasibility of OAM based wireless communications is validated, there are many research problems unsettled. For example, in order to support simultaneous transmission with multiple OAM-modes, transmit and receive antennas need to support the generation and reception, respectively, of multiple OAM-modes mixed signals. For another example, because the EM wave with OAM is vorticose hollow and divergent~\cite{Thed_2007}, the OAM beam needs to be converged for relatively long distance transmission. Moreover, the phase errors due to the non-alignment or fading are very hard to be estimated at the receiver. Although OAM beam is vorticose hollow and divergent, the divergence of OAM beams greatly reduces as the frequency increases. With small divergence, the received signal-to-noise ratio (SNR) is relatively large, which is beneficial for the reception of OAM signal. Thus, it is expected to use OAM in mmWave networks with high frequency such as 70GHz. The authors of~\cite{OAM_Wenchi_2017} proposed the framework for OAM embedded massive MIMO communication to obtain the multiplicative spectrum efficiency gain for joint OAM and massive MIMO mmWave wireless communications, which is larger than that for the traditional massive MIMO mmWave communications.

In this paper, we survey the fundamental issues of using OAM in wireless communications. We show the advantages and challenges of OAM based wireless communications. Furthermore, we give a novel OAM-modes based orthogonal multiuser access framework and evaluate the obtained spectrum efficiency in the case study.

\section{What is OAM?}
OAM is one basic physical property of EM wave. It describes the orbital property for EM rotational degree of freedom and rotation characteristic for energy. OAM is interpreted as a beam with a number of OAM-modes which can theoretically take not only any integer value but also any non-integer value. Inherently, the EM wave carried OAM can be generated by PE wave with one phase rotation factor $\exp(i l\varphi)$, where $i=\sqrt{-1}$, $l$ is the order/index of OAM-mode, and $\varphi$ is the azimuthal angle (defined as the angular position on a plane perpendicular to the axis of propagation). A pure OAM-mode is characterized by integer and different OAM-modes are orthogonal with each other. When the OAM-mode is a non-integer, the phase term $\exp(il\varphi)$ can be expressed by the sum of Fourier series of orthogonal OAM-modes. Affected by the rotation phase factor, the wavefront phase is spiral structure instead of planar structure. The wavefront phase rotates around the beam propagation direction and the phase changes $2\pi l$ after a full turn.

\begin{figure*}[h]
%\vspace{5pt}
\centering
    \subfigure[Wavefront of mode 0.]{\hspace{-0cm}
        \includegraphics[width=0.20\linewidth]{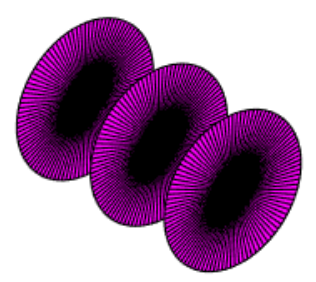}     %0.25
        \label{fig:mode0}
    }
    \subfigure[Wavefront of mode 1.]{\hspace{-0cm}
        \includegraphics[width=0.20\linewidth]{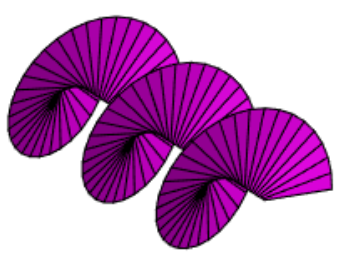}
        \label{fig:mode1}
    }
    \subfigure[Wavefront of mode 2.]{\hspace{0.5cm}
        \includegraphics[width=0.20\linewidth]{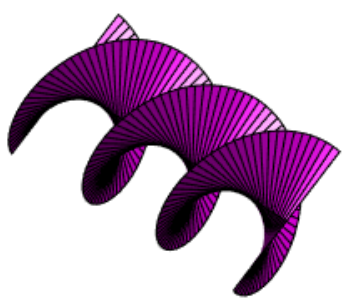}     %0.25
        \label{fig:mode2}
    }
    \subfigure[Wavefront of mode 3]{\hspace{0.5cm}
        \includegraphics[width=0.20\linewidth]{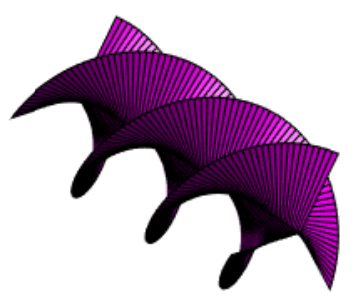}     %0.25
        \label{fig:mode3}
    }

    \subfigure[3D profile of mode 0.]{\hspace{-0.3cm}
        \includegraphics[width=0.40\linewidth]{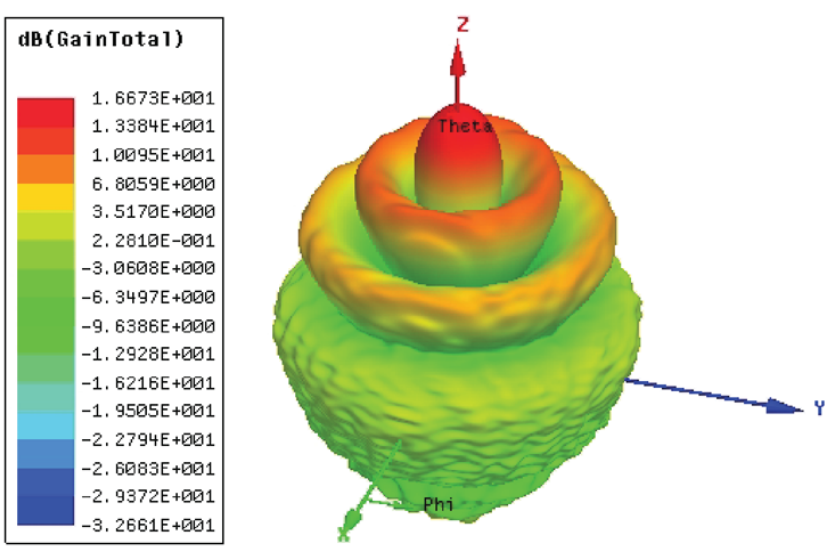}     %0.25
        \label{fig:l=0}
    }
    \subfigure[3D profile of mode 1.]{\hspace{0.5cm}
        \includegraphics[width=0.40\linewidth]{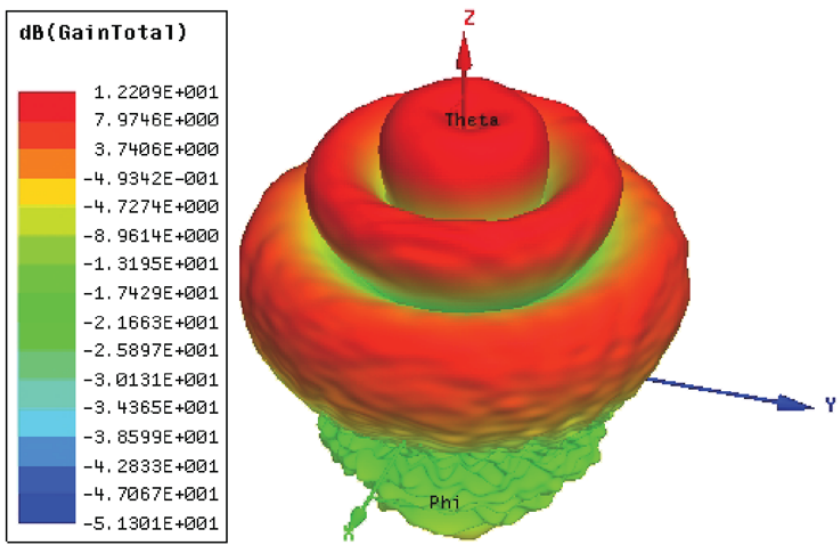}
        \label{fig:l=1}
    }
    \subfigure[3D profile of mode 2.]{\hspace{-0.2cm}
        \includegraphics[width=0.40\linewidth]{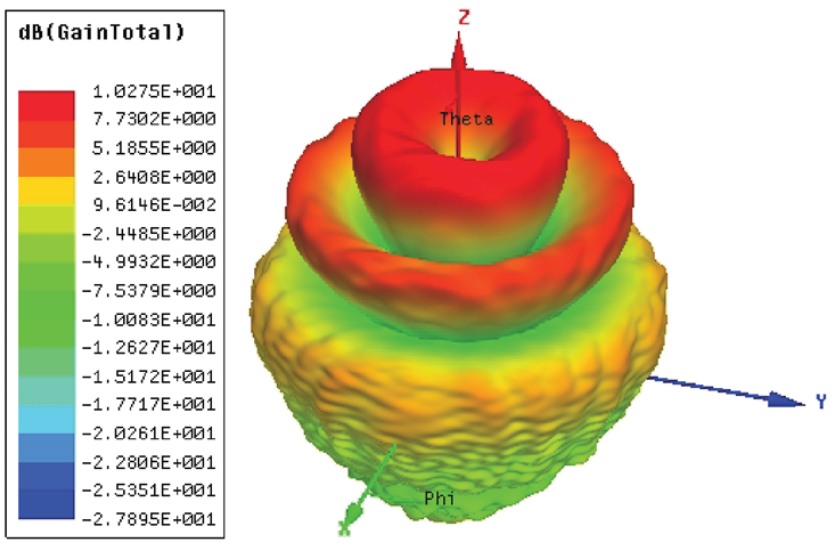}     %0.25
        \label{fig:l=2}
    }
    \subfigure[3D profile of mode 3]{\hspace{0.5cm}
        \includegraphics[width=0.40\linewidth]{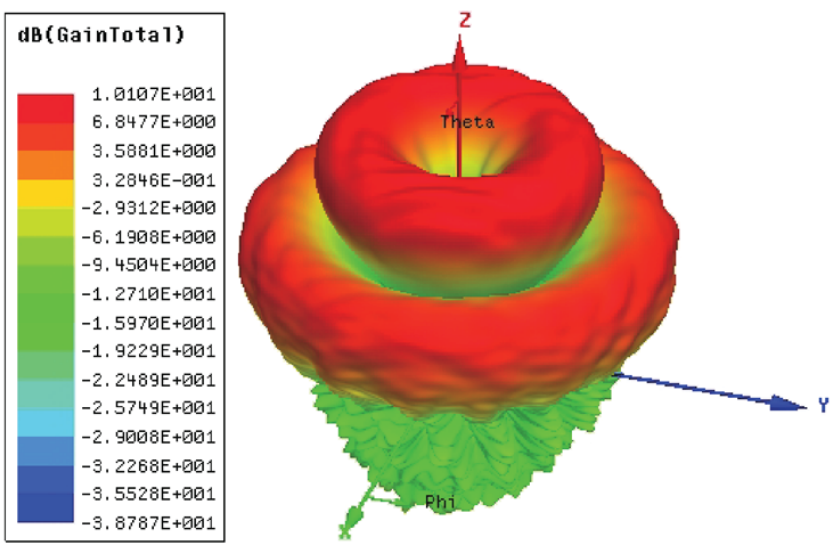}     %0.25
        \label{fig:l=3}
    }

    \caption{Wavefront and 3D profile for OAM waves with different modes.}\label{fig:wavefront}
%\vspace{-5pt}
\end{figure*}

Figure~\ref{fig:wavefront} shows the wavefront and 3 dimensional (3D) profile for OAM waves with different modes, where the transmit antenna is uniform circular array (UCA) antenna with 16 array-elements. Figs.~\ref{fig:mode0}-\ref{fig:mode3} show the wavefront phase corresponding to OAM-modes 0, 1, 2, and 3, respectively. In fact, OAM-mode 0 represents the PE wave as shown in Fig.~\ref{fig:mode0}. Based on Figs.~\ref{fig:mode0}-\ref{fig:mode3}, we can observe that the spiral characteristic of OAM wave becomes complicated and the phase changes sharply as the the order/index of OAM-modes increases within the same distance. Figs.~\ref{fig:l=0}-\ref{fig:l=3} show the 3D profiles of OAM waves for different OAM-modes 0, 1, 2, and 3, respectively. There exist central hollow for different OAM-modes except OAM-mode 0. This is because the OAM wave of mode 0 is in fact the PE wave. The central hollow increases as the order of OAM-mode increases. Also, the power gain decreases as the order of OAM-mode increases. This indicates that it is impossible for long distance OAM wave transmission by directly using OAM-modes. For long distance transmission, we need to converge the hollow OAM wave.

\section{The OAM Based Wireless Communications}
Different from frequency/time/code-domain based orthogonal division, OAM offers a new mode domain to support the orthogonal access of multiple users. With OAM, we can re-design the wireless communications because many aspects in wireless communications can be improved with the new orthogonal dimension. Three basic advantages regarding OAM based wireless communications are reviewed as follows:

{\bf Advantage 1: High spectrum efficiency} --- Different OAM-modes are orthogonal with each other. Thus, in ideal case there is no interference among different OAM-modes. With the orthogonality, the parallel transmission can be performed among multiple OAM-modes. The orthogonality among different OAM-modes can be used to increase the spectrum efficiency in wireless communications without consuming more traditional frequency/time/code/power-domain resources. Also, mode-domain resources can be jointly used with frequency/time/code-domain resources to significantly increase the spectrum efficiency in wireless communications.

%{\bf Advantage 2: High spectrum efficiency} ---- OAM, which provides a new dimension for transmission, offers a new way to solve the spectrum tension problem for wireless communications. Mode dimension and frequency dimension are mutually independent. Thus, multiple-mode transmission can be jointly used with the conventional communications to achieve high spectrum efficiency in radio vortex wireless communications.

{\bf Advantage 2: More users access} --- OAM provides a novel multiple access method, i.e., mode division multiple access (MDMA), without consuming more frequency and time resources. With MDMA, different users can employ different OAM-modes to orthogonally access the wireless networks. Instead of non-orthogonal multiple access, which uses power domain to distinguish multiple users, it is expected to get back to orthogonal multiple access using mode-domain resource in future wireless communications.

%as well as support more users

{\bf Advantage 3: High reliability for anti-jamming} --- Faced with the more and more crowded spectrum pressure, there exist limitations using the conventional frequency hopping techniques for anti-jamming. However, OAM-mode hopping technique has the potential for anti-jamming in future wireless communication. OAM can not only be used within the narrow band, but also jointly used with frequency-hopping in wide band to improve the ability of anti-jamming for wireless communications. %Thus, OAM used for anti-jamming can jointly be used with the conventional frequency hopping to further guarantee the reliability requirement of system.

Although OAM has the potential to increase the spectrum efficiency, support more users, and improve the reliability of anti-jamming, there are still some important research challenges remaining unsettled. The issues can be classified into three categories: radio vortex signal generation, transmission, and reception.

\begin{figure*}[t]
%\vspace{5pt}
\centering
    \subfigure[Spiral Phase Plate (SPP) antenna.]{\hspace{-0cm}
        \includegraphics[width=0.3\linewidth]{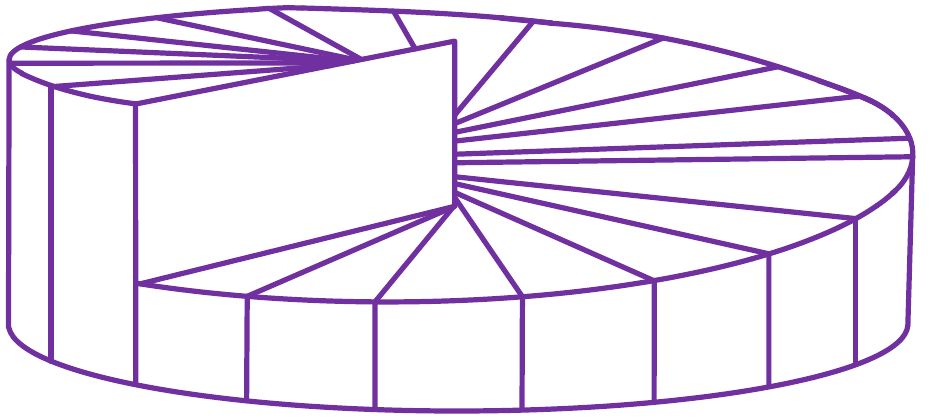}     %0.25
        \label{fig:SPP}
    }
    \subfigure[Uniform Circular Array (UCA) antenna.]{\hspace{-0cm}
        \includegraphics[width=0.23\linewidth]{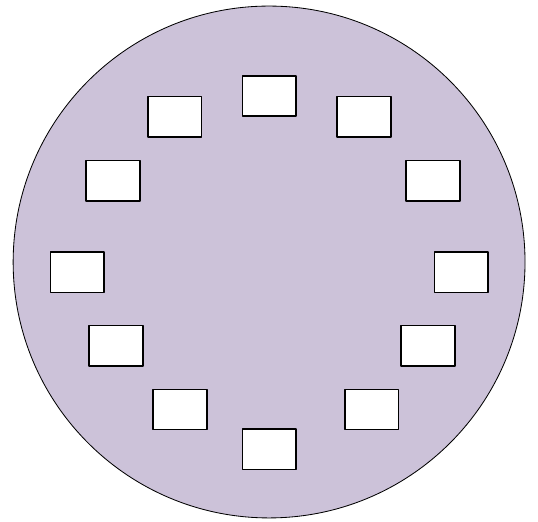}
        \label{fig:UCA}
    }
    \subfigure[Metasurface.]{\hspace{0.5cm}
        \includegraphics[width=0.3\linewidth]{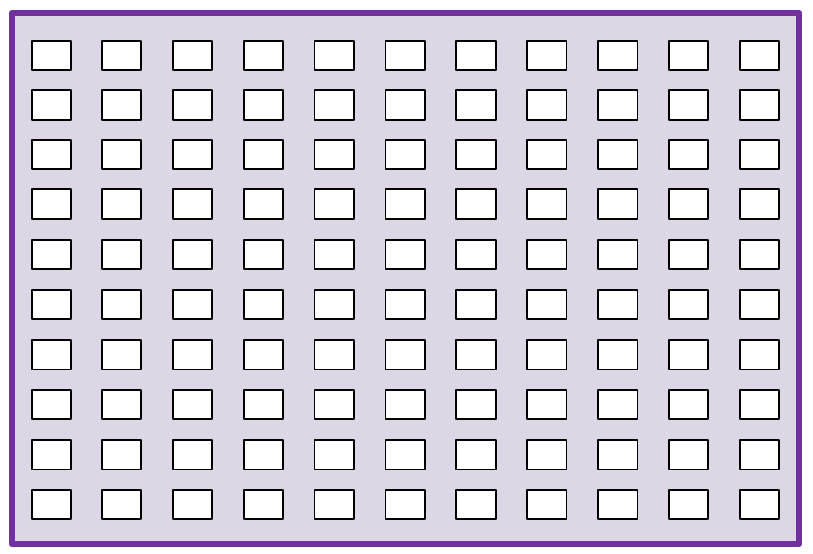}     %0.25
        \label{fig:meta}
    }

    \caption{Three kinds of antenna structures for radio vortex signal generation.}\label{fig:antenna}
%\vspace{-5pt}
\end{figure*}

{\bf Radio Vortex Signal Generation.}
Different from traditional PE wave based signals, radio vortex signals have the phase rotation factor $\exp(i l\varphi)$. There are some popular facilities can be used to generate radio vortex signal such as Spiral Phase Plate (SPP) antenna, Uniform Circular Array (UCA) antenna, and metasurfaces, as shown in Fig.~\ref{fig:antenna}.

\begin{itemize}
  \item SPP antenna~\cite{LOS_OAM_2017}: An example of SPP antenna is given in Fig.~\ref{fig:SPP}. The SPP antenna generates the phase delay by increasing the antenna thickness in proportion to the azimuthal angle or by drilling inhomogeneous holes in dielectric plate to change the equivalent permittivity. The SPP antenna has the advantages of small divergence and low attenuation as well as the disadvantages of not applicable for relatively low frequency transmission and cannot generate multiple OAM-modes simultaneously.
  \item UCA antenna~\cite{Gaoyue_TAP_201707}: An example of UCA antenna is given in Fig.~\ref{fig:UCA}. The phase information of adjacent array-element of UCA antenna is linearly increased by $2\pi l/N$, where $N$ is the number of array-elements. The UCA antennas are low profile, low weight, and easy to manufacture with rectangular patch arrays. Also, the UCA antennas can simultaneously generate multiple vortex beams with multiple OAM-modes even in the radio frequency band. However, the vortex beams generated by UCA is divergent and centrally hollow. Thus, the UCA antennas need to be jointly used with the converging schemes to combat the signal attenuation during the propagation.
  \item Metasurfaces\cite{Yu2016Design}: An example of metasurfaces is given in Fig.~\ref{fig:meta}. In the metasurfaces based OAM signal generation schemes, the wavefront of electromagnetic waves are controlled by regulating phase shift to the incoming waves. These schemes have the advantages of low profile, small mass, and low manufacturing cost. However, it is hard to accurately control the phase for signal modulation and thus not applicable to multiple OAM-modes transmission in wireless communications.
\end{itemize}

%Based on the advantages and disadvantages of the above three type of antennas, we recommend the UCA antenna for wireless communications mainly due to the property that it is convenient to generate multiple OAM-modes with one transmit UCA antenna.

{\bf Radio Vortex Signal Transmission.}
Three typical problems need to be considered for radio vortex signal transmission.
%1. Alignment
%2. Fading
%3. Convergence

\begin{itemize}
\item \textit{Transmitter-receiver alignment.} It is required that the transmitter and receiver are aligned with each other to decompose signals with different OAM-modes~\cite{concentric_ICCC2017_OAM}. If the transmitter and the receiver are not aligned, the phase of received signal contains not only the phase of OAM mode, but also the phase turbulence due to unequal distance transmission at different places of the receiver. Therefore, for non-aligned scenarios, it is demanded to add phase turbulence adaptive estimation algorithm at the receiver.
    %If they are coaxial but not parallel, the phases of OAM-modes are different because of the distance difference caused by the misalignment between the transmitter and receiver. It is crucial to design algorithms to demultiplex the signals with different OAM-modes. If they are non-coaxial, it is possible that the receiver cannot detect the transmit signals because the receive phases of OAM-modes are different with the transmit phases of OAM-modes. It is important to distinguish the OAM-modes by proposing new methods.
  \item \textit{Fading.} Current OAM related researches mainly focus on the line-of-sight scenario, where no fading has been taken into account. However, there exists fading in many practical scenarios, leading to the randomness of wavefront phases at the receiver. If the signals of different OAM-modes undergo fading, the phase change corresponding to each OAM-mode needs to be estimated.
  \item \textit{Convergence.} Because an OAM beam becomes more and more divergent as the order of OAM-mode increases, it severely reduces the transmission distance and decreases the spectrum efficiency of OAM based wireless communications. It is required to make OAM beams convergent so that all OAM-modes including both high and low order OAM-modes can be efficiently used. However, it is a difficult task to design efficient antenna structure and algorithms to converge the OAM beams without changing the original wavefront phases of OAM-modes. There are two typical methods for converging OAM beams: parabolic antenna and lens antenna. The parabolic antenna reforms the diverged OAM beam to an converged beam while keeping remain the angular identification information of each OAM-mode. The parabolic antenna enjoys relatively low attenuation for OAM beams. However, the size of the parabolic antenna is relatively large. Through refracting, the lens antenna can reform the diverged OAM beams into the converged OAM beams. The lens antenna fits all frequency band. However, it is bulky and needs to endure relatively large attenuation.

\begin{figure*}%[h]
    \centering
    \vspace{5pt}

    \subfigure[The E field of unconverged OAM beams.]{
\centering
\begin{minipage}[t]{0.22\linewidth}
\centering
\centerline{\includegraphics[width=3.2cm,height=5.5cm]{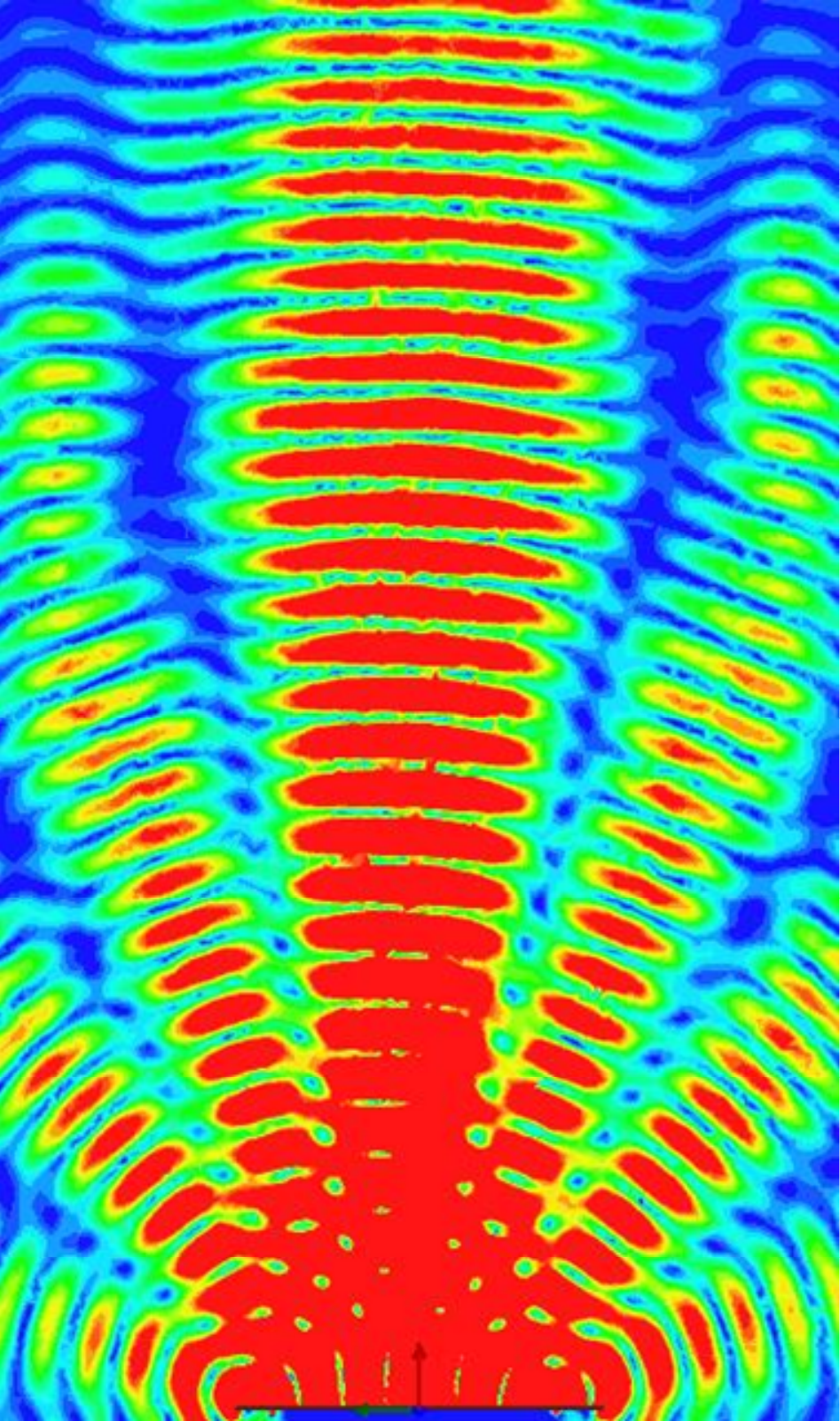}}
\centerline{$l = 0$ (PE beam)}
\end{minipage}
\centering
\begin{minipage}[t]{0.22\linewidth}
\centering
\centerline{\includegraphics[width=3.2cm,height=5.5cm]{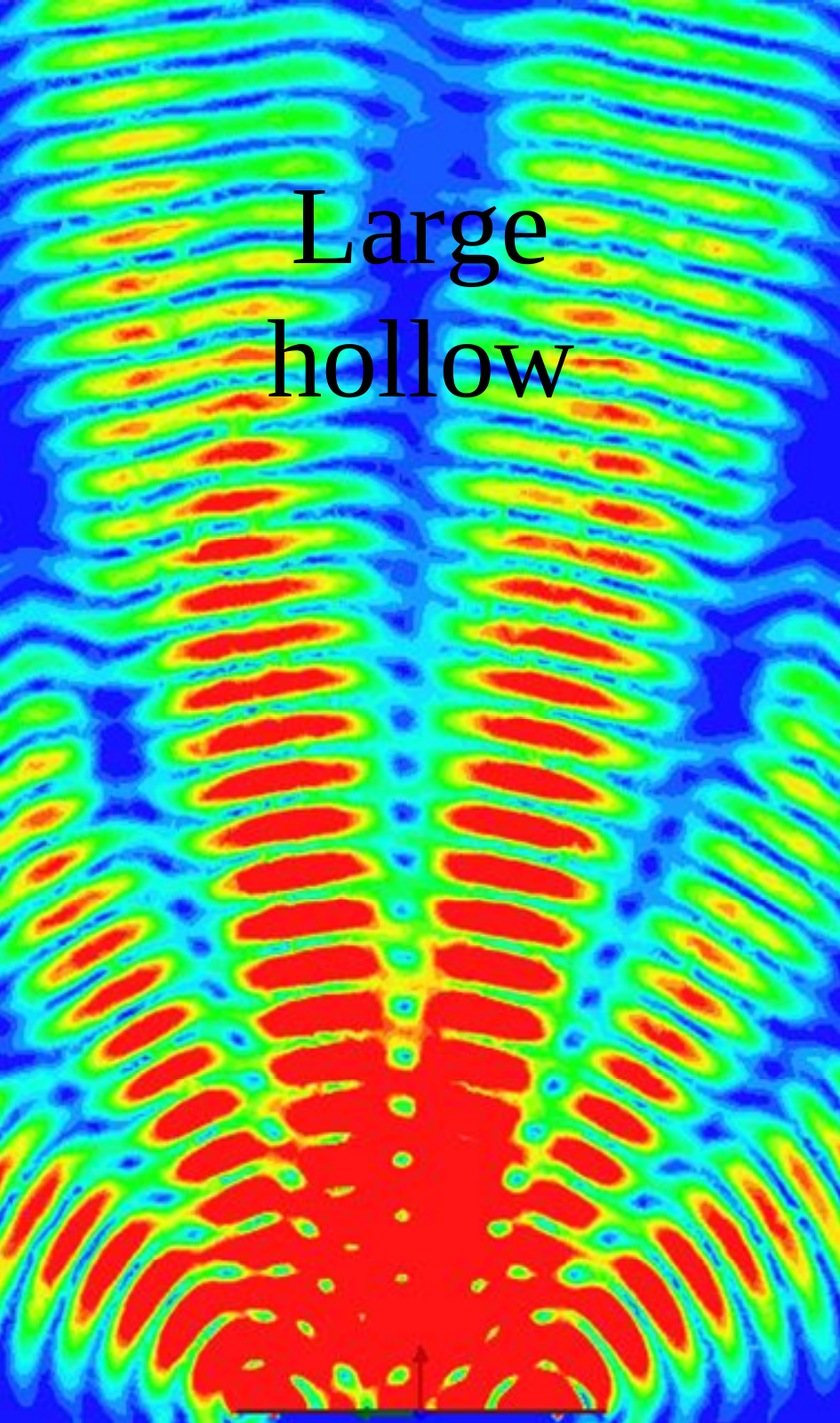}}
\centerline{$l = 1$}
\end{minipage}
\centering
\begin{minipage}[t]{0.22\linewidth}
\centering
\centerline{\includegraphics[width=3.2cm,height=5.5cm]{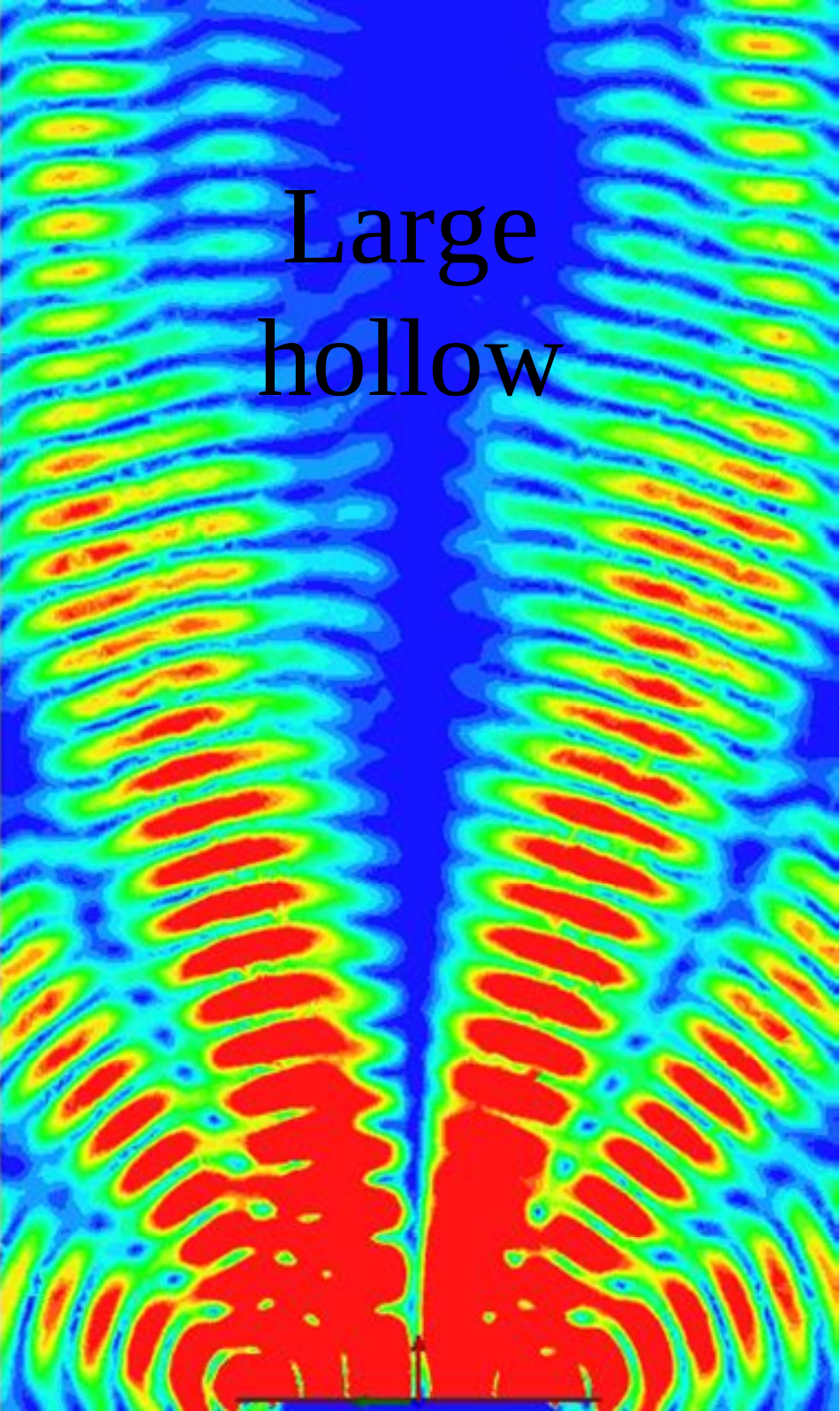}}
\centerline{$l = 2$}
\end{minipage}
\centering
\begin{minipage}[t]{0.22\linewidth}
\centering
\centerline{\includegraphics[width=3.2cm,height=5.5cm]{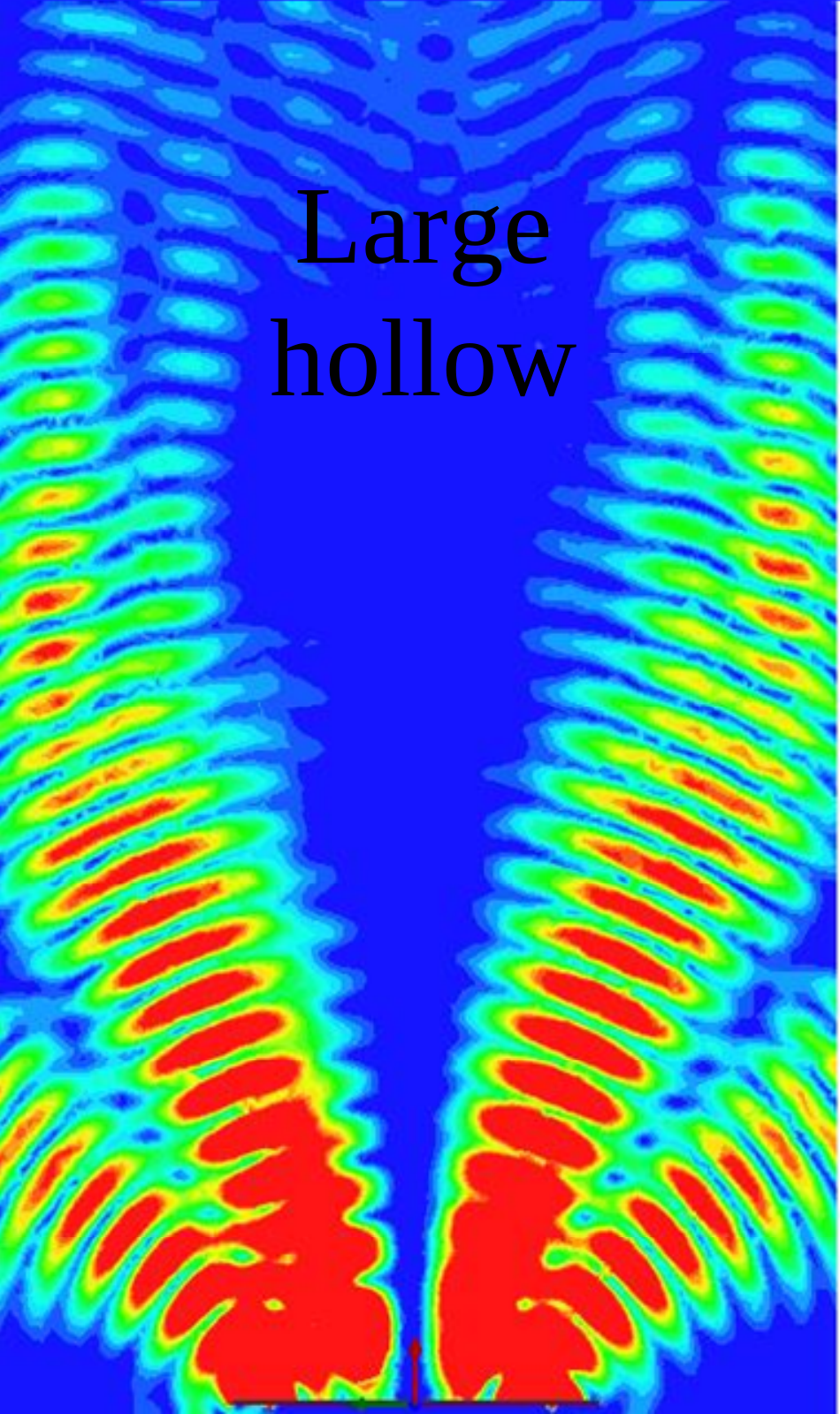}}
\centerline{$l = 3$}
\end{minipage}
    }

    \subfigure[The E field of lens converged OAM beams.]{
\centering
\begin{minipage}[t]{0.22\linewidth}
\centering
\centerline{\includegraphics[width=3.2cm,height=5.5cm]{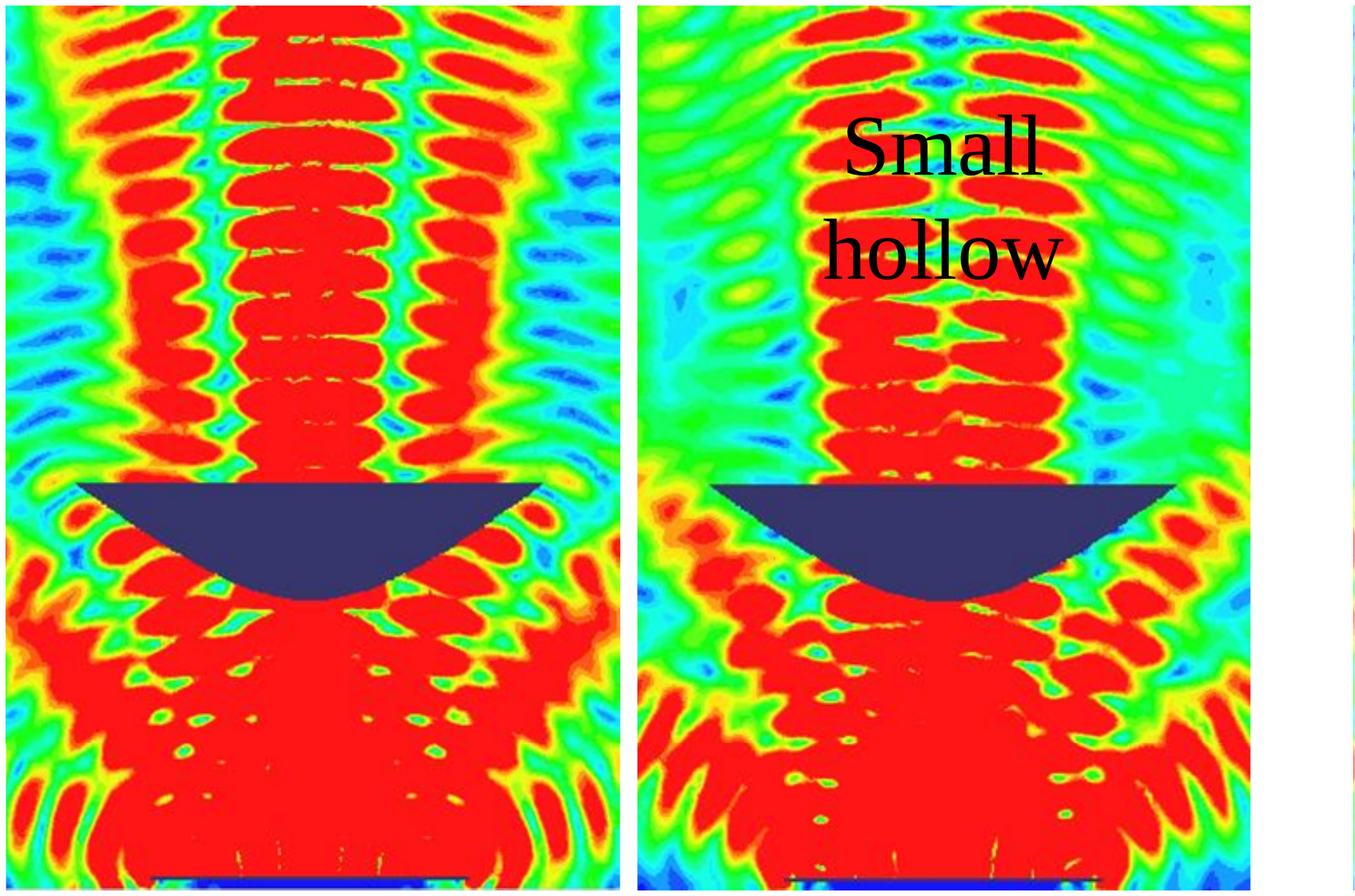}}
\centerline{$l = 0$ (PE beam)}
\end{minipage}
\begin{minipage}[t]{0.22\linewidth}
\centering
\centerline{\includegraphics[width=3.2cm,height=5.5cm]{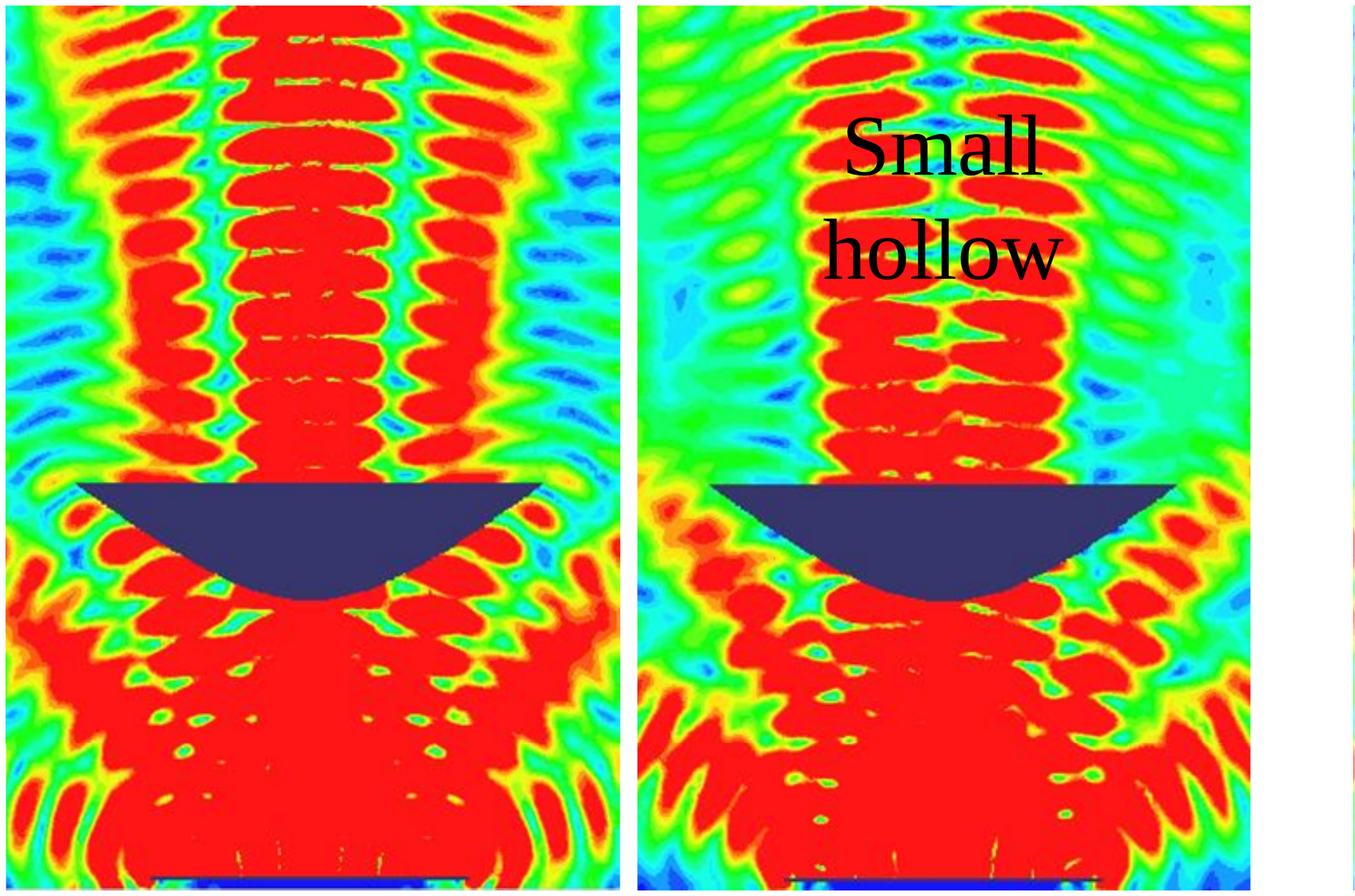}}
\centerline{$l = 1$}
\end{minipage}
\centering
\begin{minipage}[t]{0.22\linewidth}
\centering
\centerline{\includegraphics[width=3.2cm,height=5.5cm]{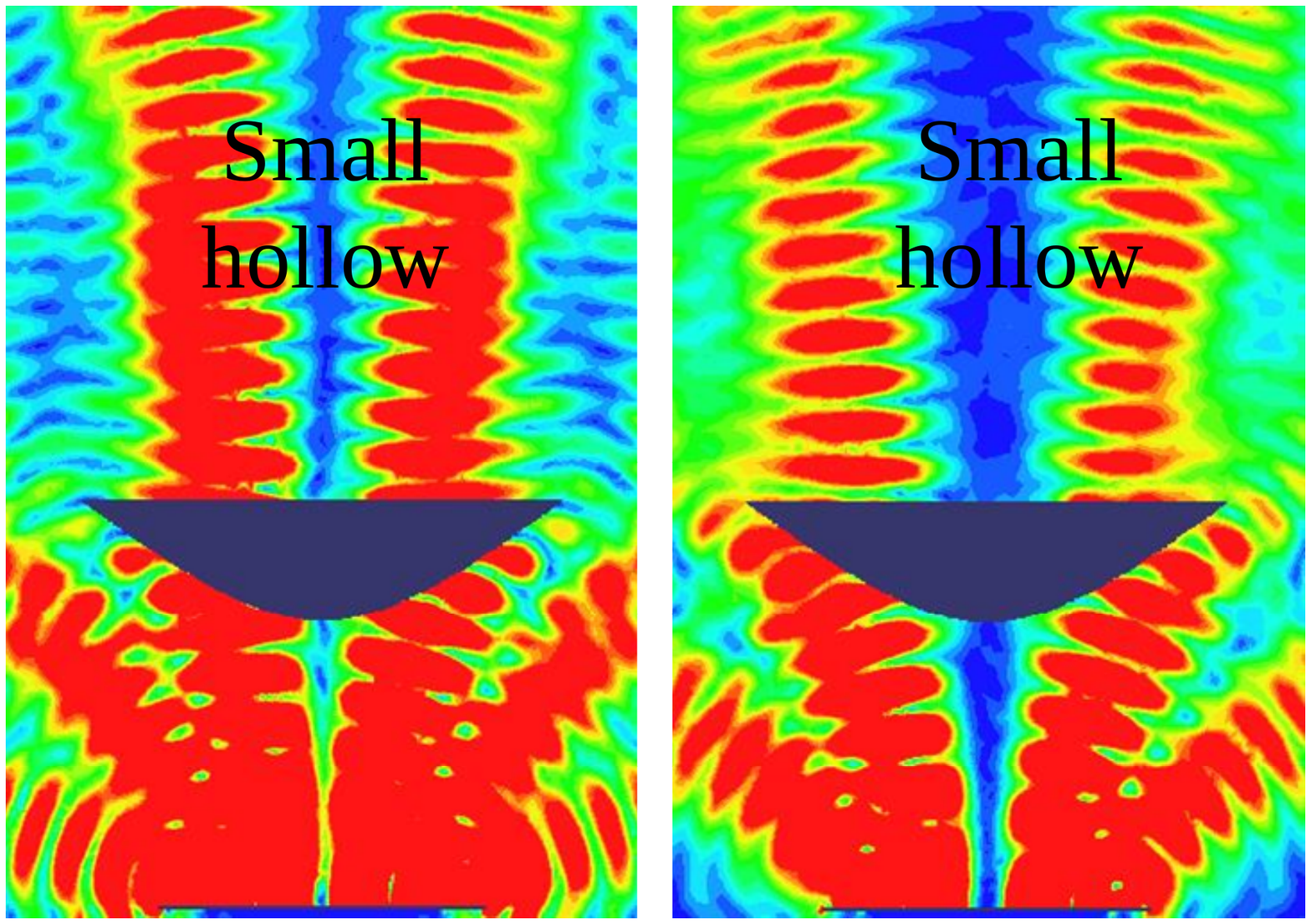}}
\centerline{$l = 2$}
\end{minipage}
\centering
\begin{minipage}[t]{0.22\linewidth}
\centering
\centerline{\includegraphics[width=3.2cm,height=5.5cm]{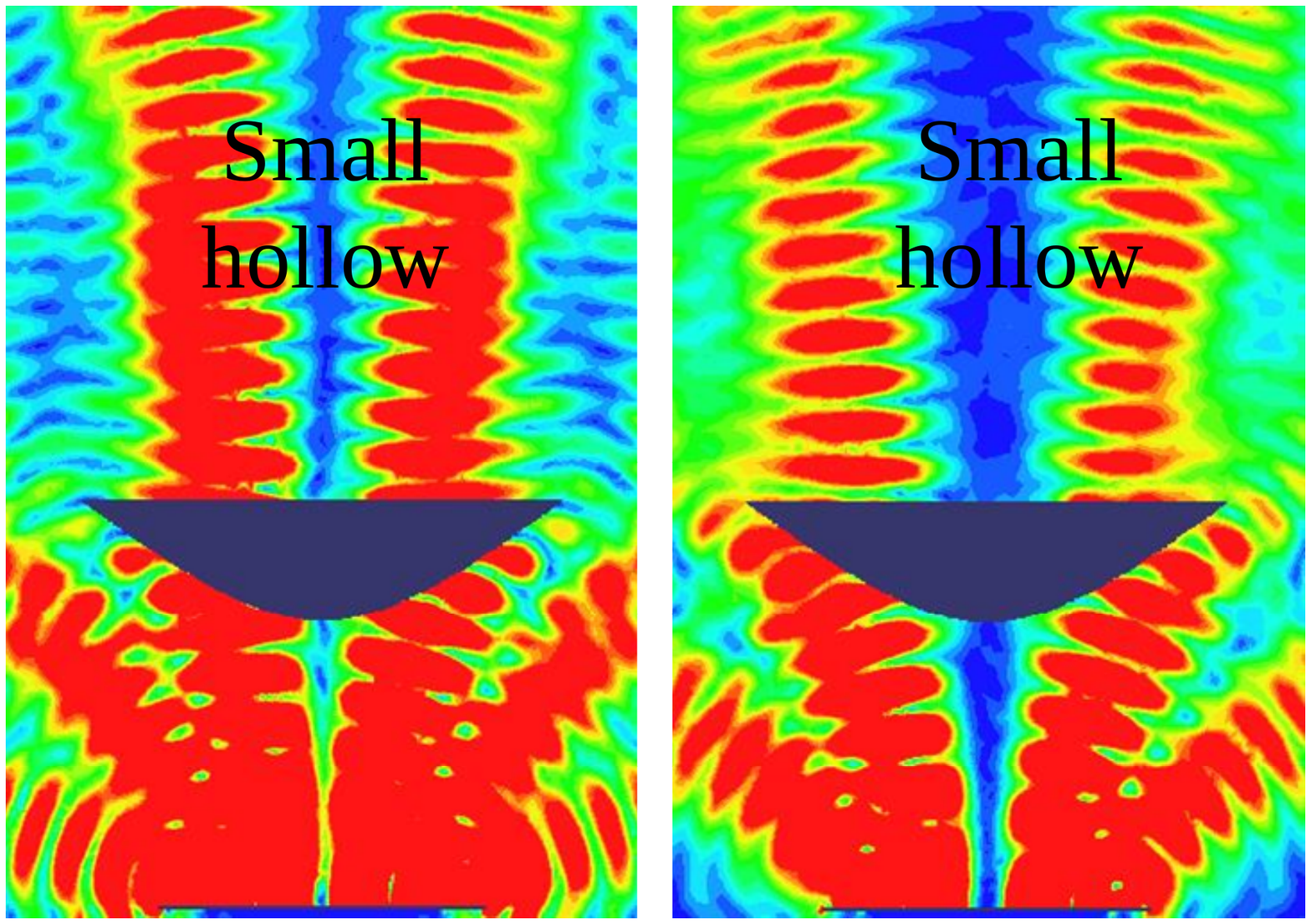}}
\centerline{$l = 3$}
\end{minipage}
    }
    \subfigure[The phase profiles of OAM-mode 0, 1, 2, and 3.]{
\centering
\begin{minipage}[t]{0.22\linewidth}
\centering
\centerline{\includegraphics[width=0.9\linewidth]{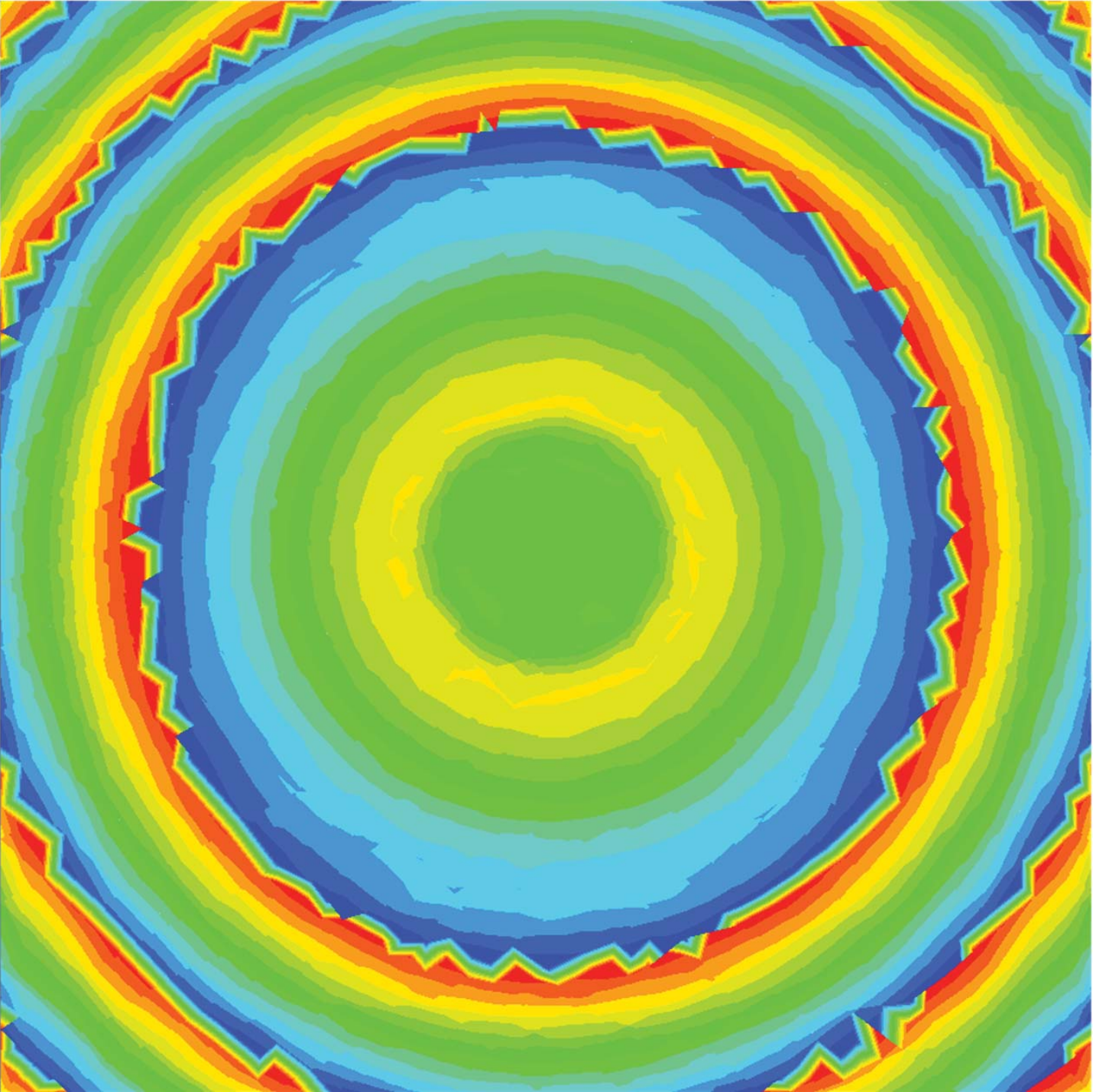}}
\centerline{$l = 0$ (PE beam)}
\end{minipage}
\centering
\begin{minipage}[t]{0.22\linewidth}
\centering
\centerline{\includegraphics[width=0.9\linewidth]{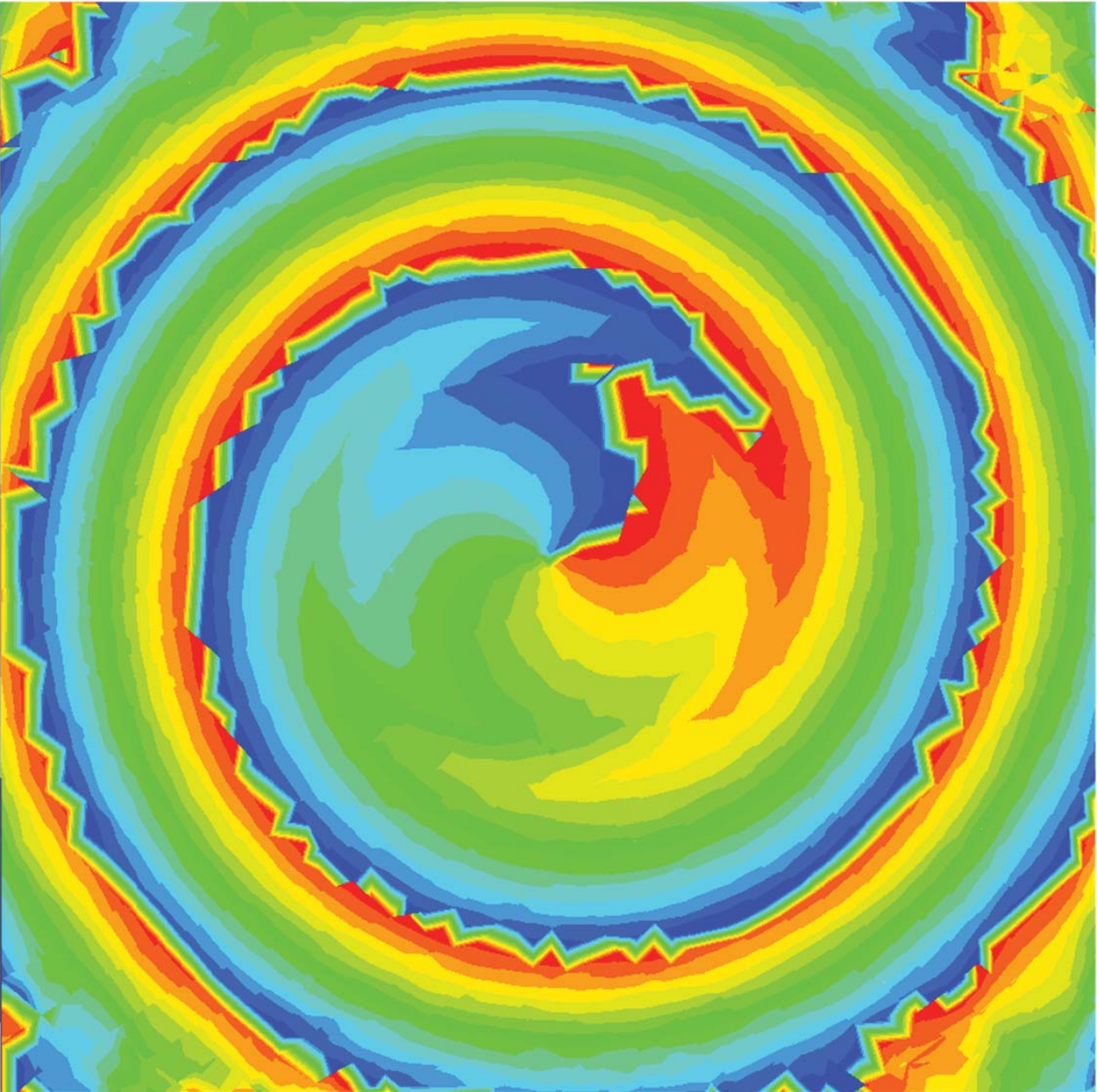}}
\centerline{$l = 1$}
\end{minipage}
\centering
\begin{minipage}[t]{0.22\linewidth}
\centering
\centerline{\includegraphics[width=0.9\linewidth]{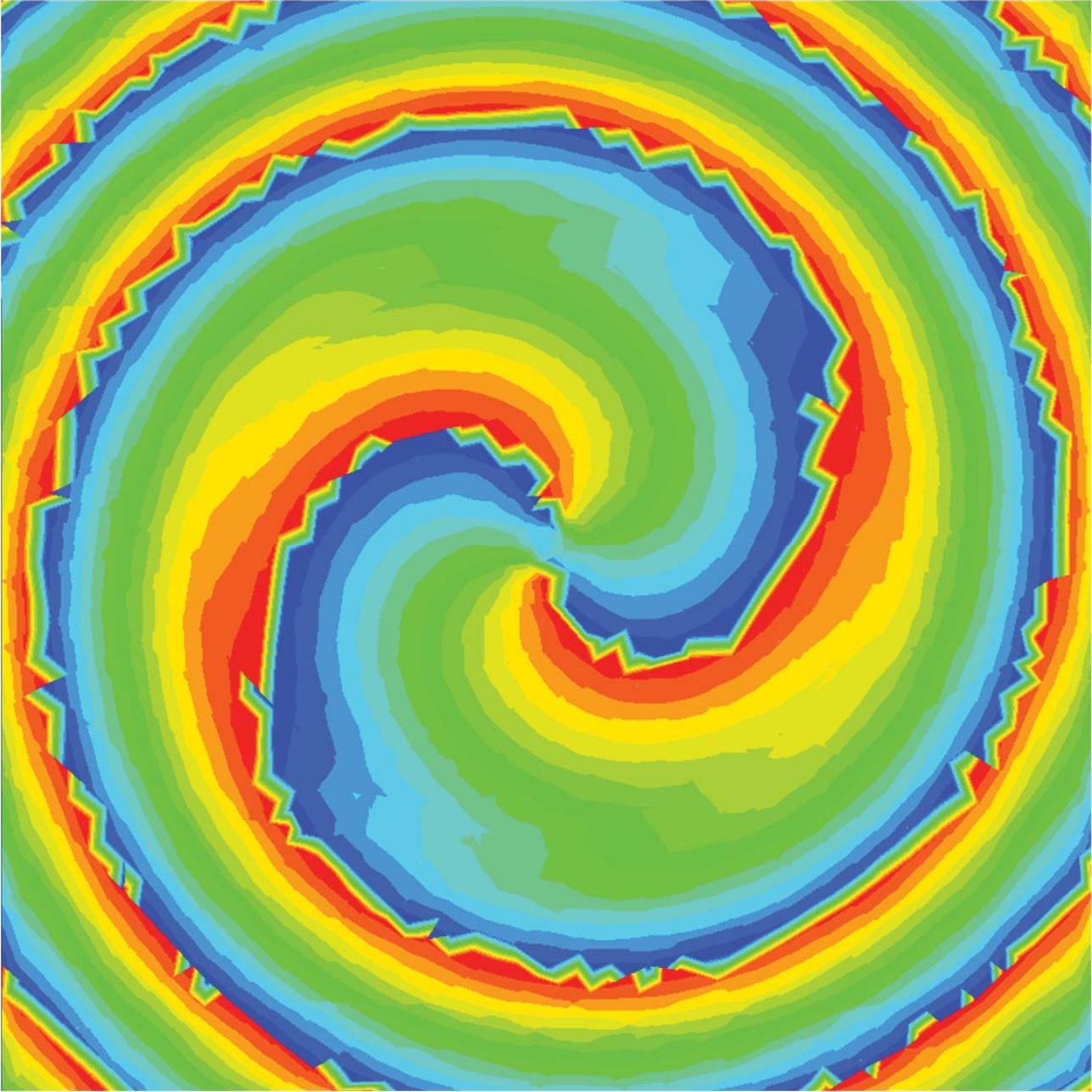}}
\centerline{$l = 2$}
\end{minipage}
\centering
\begin{minipage}[t]{0.22\linewidth}
\centering
\centerline{\includegraphics[width=0.9\linewidth]{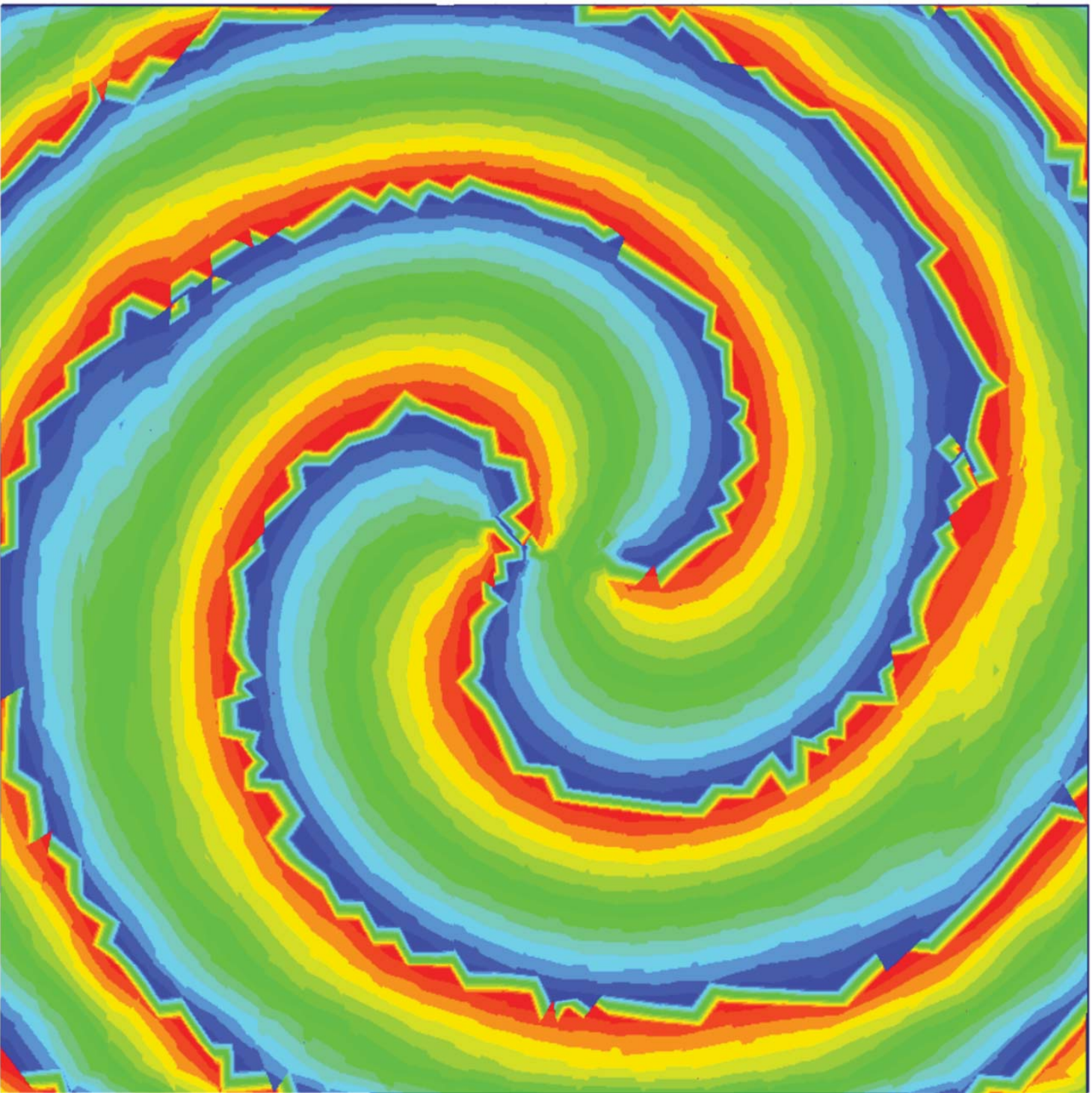}}
\centerline{$l = 3$}
\end{minipage}
  }
    \caption{The E field of unconverged and converged OAM beams observed from the horizontal direction and the phase profiles of the OAM-modes.}
    \label{fig:E_field}
    \vspace{-10pt}
\end{figure*}

  \ \ Figure~\ref{fig:E_field} shows the E field of unconverged and converged OAM beams observed from the horizontal direction and the phase profiles of the OAM-modes. $N$ patch-elements of UCA are equally distributed on the circle, where the radius is set as 25~mm. The relative permittivity of patch material is set as 2.2. The length and the width of patch element are set as 2.947~mm and 3.388~mm, respectively. Then, we place the plane at 100 mm to observe the results. Figs.~\ref{fig:E_field}(a) and~\ref{fig:E_field}(b) show the E field of unconverged and converged OAM beams, respectively, observed from the horizontal direction. Fig.~\ref{fig:E_field}(a) confirms that the unconverged OAM beams have a large hollow in the central area. Moreover, the hollow increases as the order of OAM-mode ($l$) increases. Fig.~\ref{fig:E_field}(b) verifies that the OAM beams can be effectively converged with the lens antenna. The central hollow is very small as compared with Fig.~\ref{fig:E_field}(a) and the intensity of OAM beams efficiently increases. Fig.~3(c) shows the phase profiles of the OAM-modes 0, 1, 2, and 3. We set the frequency of OAM beams as 35 GHz.
 % \item Transmitter-receiver alignment.
%
%  different distance --> transmission phase different --> the detection algorithm depends on the rotation of the UCA receive antenna
%
%
%  \item Fading. Current  applied to XXX (LOS). However, fading leads to the randomness of wavefront wave. If want to use OAM in fading channel, the exact phase change needs to be estimated.
%
%  \item Convergence
  %\item Encoding~\cite{GC17_OAM}. The encoding method for OAM beams uses Huffman coding. The Huffman coding based adaptive mode modulation scheme can adaptively choose the OAM modes to increase the spectrum efficiency of OAM based radio vortex wireless communications.
  %\item The metamaterials can change the angular distribution of electromagnetic waves to generate the OAM signals. It is easy to implement the metamaterials based OAM beams converging.
\end{itemize}

{\bf Radio Vortex Signal Reception.}
At the receiver, the phase detection is a key to distinguish the order of different OAM-modes~\cite{Uchida2010Generation}. The radio vortex signal reception schemes for SPP antenna, UCA antenna, as well as other schemes are discussed in the following. %The OAM beams are required to be coaxial, otherwise, it is possible for non-coaxial OAM beams to interfere with each other.
\begin{itemize}
\item Reception with SPP antenna. One SPP antenna can detect the signal with one OAM-mode. The receiver generates the conjugate phase related to the corresponding OAM-mode carried by the signal. Then, multiplied with the received one OAM-mode signal, the OAM wave can be transformed into plane wave. An SPP antenna can only decompose one OAM-mode among different OAM-modes due to the characteristic of SPP antenna~\cite{LOS_OAM_2017}.

\item Reception with UCA antenna. The spatial fast Fourier transform (FFT) can be used to obtain the signal carried by the corresponding OAM-mode~\cite{OAM_Wenchi_2017}. This is because it has a property that after spatially sampling the sum is zero within the interval length except the designed OAM-mode. The UCA antenna based reception scheme supports the detection for multiple OAM-modes. However, it is highly required that the transmit UCA and the receive UCA are aligned with each other.

\item Other schemes. There exist other schemes to obtain the signals carried by OAM-modes. The phase gradient method (PGM) is developed to identify the OAM-modes. The PGM is dependent on the circular separation among receiver antennas. Because the helical phase fronts of different OAM beams are different, the OAM-mode can be detected using a two-point phase measurement. Other reception schemes include direct torque measurement and triangulation.
\end{itemize}

\section{OAM-Modes Based Multiple Access: A Case Study}
\begin{figure}
\centering
\includegraphics[scale=0.3]{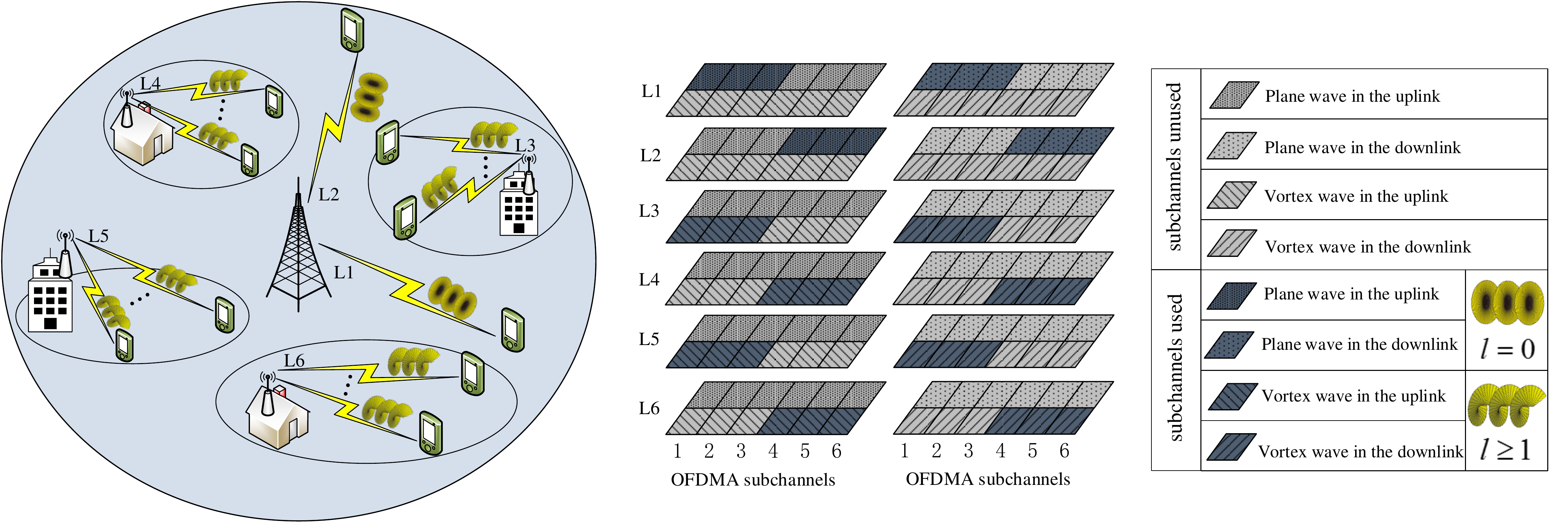}
\caption{The OAM beams based multiple users access in wireless networks.} \label{fig:UDN_interferenceOAM}
%\vspace{-0.2cm}
\end{figure}

As a case study, we investigate the OAM-modes based multiple access in wireless networks, as shown in Fig.~\ref{fig:UDN_interferenceOAM}, consisting one macrocell and several small cells. $L1$-$L2$ denote the PE beam used in macrocell and $L3$-$L6$ represent the OAM beams corresponding to a group of OAM-modes used in small cells.
%Since the number of small cells in HetUDNs is extremely large, different kinds of interference co-exist in HetUDNs, which makes the interference management very hard. Another reason why the interference is very difficult to avoid is that the resource used now such as frequency and time is very limited. The wireless communications~\cite{Willner_OAM}, which can provide the new angular degree of freedom, has been paid much attention recently and can be efficiently applied into HetUDNs.
Theoretically, the cross-layer and co-layer interference can be solved with sufficiently large number of OAM-modes in wireless networks. The macrocells and small cells can utilize the PE wave and OAM waves, respectively, without causing cross-layer interference between macrocells and small cells. Also, different small cells can use different OAM-modes so that in principle there is no interference among small cells if the number of available OAM-modes is larger than the number of small cells. If not, we can develop the OAM-modes allocation/scheduling schemes for small cells to minimize the interference. For instance, the small cells far away use the same OAM-modes while the neighbouring small cells utilize different OAM-modes. As a result, the spectrum efficiency of wireless networks can be significantly increased to meet the demand of tremendous data traffic.

\begin{figure}
\centering
\includegraphics[scale=0.75]{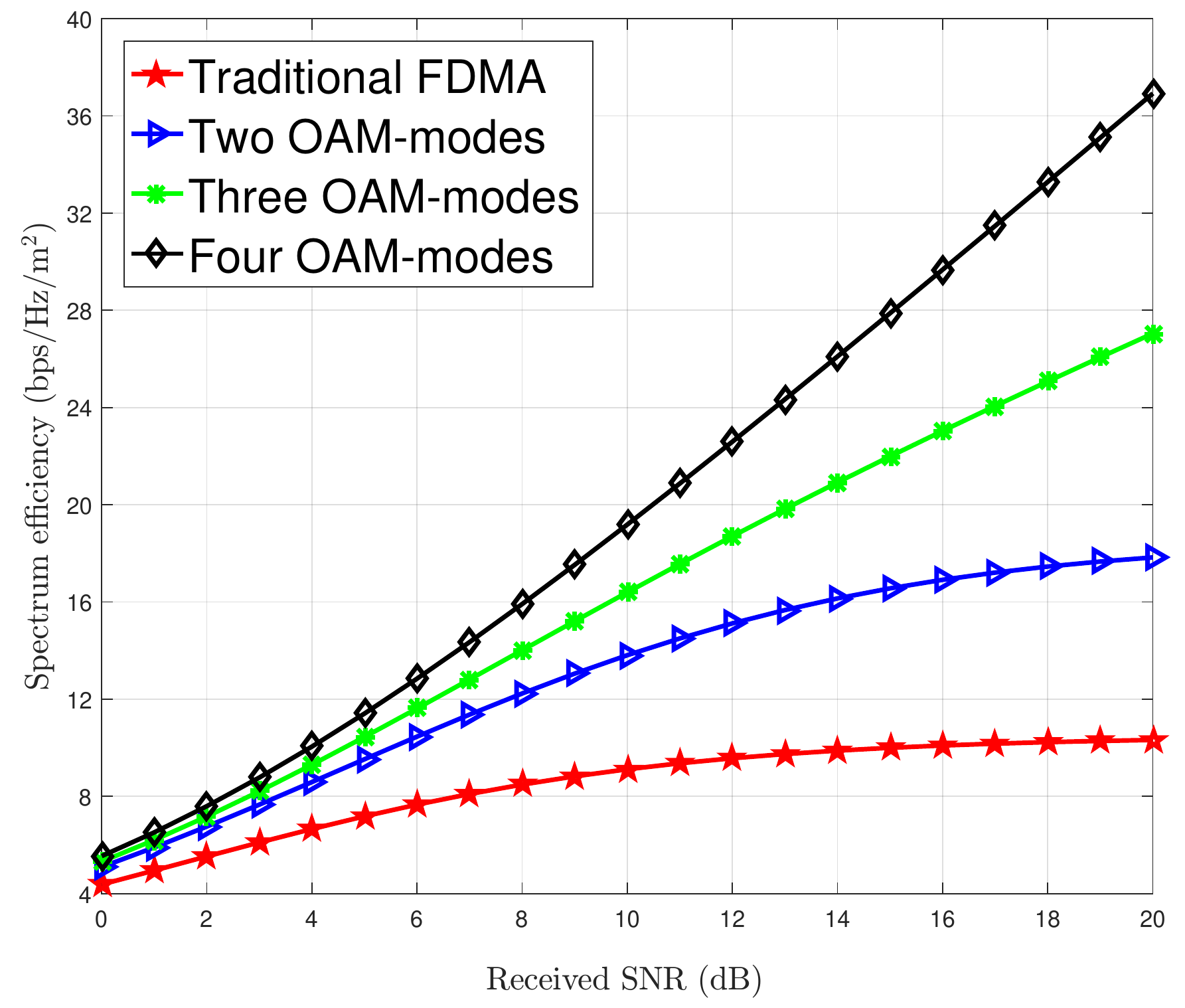}
\caption{The spectrum efficiencies versus received SNR with different number of OAM-modes for multiple users access.} \label{fig:capacity_1}
%\vspace{-10pt}
\end{figure}

We consider the 3GPP Rel-12 small cells scenarios for performance evaluation. Figure~\ref{fig:capacity_1} shows the comparison in spectrum efficiencies of traditional frequency-division-multiple-access (FDMA) and the newly proposed MDMA with different number of OAM-modes, where we set the density of users as 0.2 users/m$^2$ and the number of frequency-orthogonal channels is two. From Fig.~\ref{fig:capacity_1}, we can observe that the spectrum efficiency of more than one OAM-modes is larger than that of traditional FDMA and the spectrum efficiency significantly increases as the number of OAM-modes increases. This is because the orthogonality of OAM-modes makes the interference partly avoided. When the SNR is relatively large, there exists the ceiling effect for the spectrum efficiency. This is because the interference caused by other users increases as the SNR increases.

\begin{figure}
\centering
\includegraphics[scale=0.75]{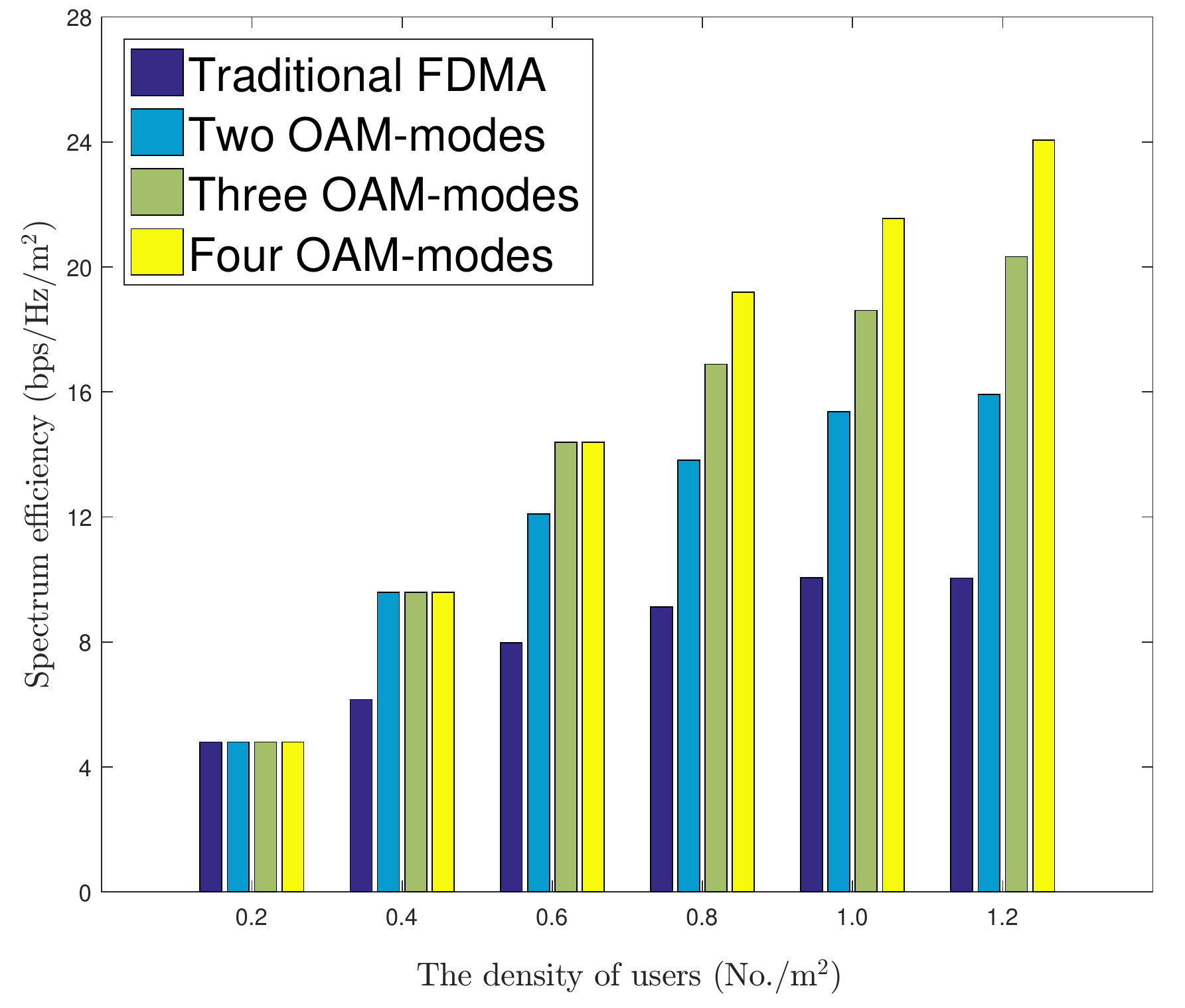}
\caption{The spectrum efficiencies versus the density of users with different number of OAM-modes for multiple users access.} \label{fig:user}
%\vspace{-10pt}
\end{figure}

Figure~\ref{fig:user} depicts the spectrum efficiencies of traditional FDMA and different number of OAM-modes, respectively, versus the density of users in wireless networks. The spectrum efficiency increases as the density of users increases in wireless networks. We can also observe that the spectrum efficiency of traditional FDMA is smaller than that of more than one OAM-modes. However, for one OAM-mode, the spectrum efficiency with 1.0 users/m$^{2}$ is similar to the spectrum efficiency with 1.2 users/m$^{2}$. When the density of users is relatively large, the spectrum efficiency of four OAM-modes is much higher than that of one OAM-mode. This is because as the number of OAM-modes increases, more users can be interference-free. Therefore, it is efficient for interference avoidance when applying wireless communications into wireless networks, thus greatly increasing the spectrum efficiency of wireless networks.

%Figure~\ref{fig:UDN_interferenceOAM} gives the system model for OAM beams based HetUDNs, where $L1$-$L2$ denote the PE beam used in macrocell and $L3$-$L6$ represent the OAM beams corresponding to a group of OAM-modes used in small cells. The macrocell uses the PE beam, i.e., the OAM-mode with $l = 0$, for data transmission while small cells utilize vortex beams with OAM-modes $l\geq 1$. Thus, the cross-layer interference can be theoretically avoided in HetUDNs. Also, different small cells employ different OAM-modes for their data transmission. Thus, the co-layer interference among different small cells can be principally avoided.

%OAM-UDN优点

%Once wireless communications is brought into HetUDNs, the interference can theoretically be avoided. As a result, the spectrum efficiency of HetUDNs can be significantly increased to meet the demand of heterogeneous tremendous data transmission~\cite{Me_Network}. Also, other performance such as power efficiency and quality of service can be improved~\cite{Me_JSAC_201612}.

However, there are some challenges regarding applying OAM beams into wireless networks.

{\bf Challenge 1: Limited Number of Available OAM-Modes.} --- The OAM beams of high order OAM-modes experience severe attenuations in propagation. Thus, the number of available OAM-modes is relatively small. Considering the high density of small cells, it is very likely the number of available OAM-modes is not enough. To significantly increase the number of available OAM-modes, the OAM beams corresponding to high order OAM-modes need to be converged, which remains as a challenging problem. Nowadays, some schemes have been developed to converge OAM beams. For example, the parabolic antenna can be used to reduce the divergence of OAM beams without impacting the orthogonality of OAM-modes~\cite{OAM_Wenchi_2017}. In addition, bifocal lens antenna also has good performance for converging OAM beams. Both of these two methods need an antenna to generate OAM beams and another antenna for converging. Alternatively, by transforming the singular UCA into the concentric UCAs, more low-order OAM-modes can be used as well as taking into account the inter-mode interference~\cite{concentric_ICCC2017_OAM}.

{\bf Challenge 2: Joint OAM-Mode and Frequency/Time Partition.} --- When the number of OAM-modes is not enough for wireless networks even after OAM-beams converging, jointly using OAM-modes and frequency/time partition are highly needed to balance the traffic over mode, frequency, and time domains to increase the spectrum efficiency~\cite{OFDM_GLOBE_2016}. Moreover, when OAM-modes and other resources (frequency or time) are combined to maximize the spectrum efficiency, the priority of choosing OAM-modes or other resources should be considered. Mode-division-multiple-access based multi-channel medium access control (MAC) protocols can be jointly integrated with frequency-division-multiple-access/time-division-multiple-access/code-division-multiple-access for high throughput and flexible multiple access in OAM wireless communications networks.

{\bf Challenge 3: Channel Estimation for Different OAM-Modes.} --- The number of channels corresponding to all OAM-modes is very large, thus causing heave overhead for channel estimation. Moreover, it is possible that the phase front of OAM-modes will change during the propagation of OAM beams, thus causing the loss of phase identification feature at the receiver. As a result, it is difficult to recover the transmit signals corresponding to different OAM-modes when the number of channels is relatively large. Until now, this area lacks very intriguing or solid results. For UCA antennas, the channel matrix for array-elements between the transmitter and the receiver is the circular matrix, which makes the channel estimation easy to handle and is likely to be the breakthrough for channel estimation in OAM wireless communications.

\section{Conclusions}\label{sec:conc}
In this article, we have introduced the fundamental theory of OAM and the OAM based wireless communications. The benefits and challenges in OAM based wireless communications are discussed. Further, we proposed a new multiuser access based on OAM beams and then we studied the case in two-tier wireless networks. The performance results verified the convergence for OAM beams and showed the spectrum efficiency enhancement for wireless networks. In conclusion, OAM provides the new mode domain, which gives many opportunities for future wireless communications research.

\end{document}